\documentclass[conference]{IEEEtran}
\ifCLASSINFOpdf

\else

\fi

 \makeatletter

\newcommand{\Rmnum}[1]{\expandafter\@slowromancap\romannumeral #1@}
\makeatother
\usepackage{cleveref}
\usepackage{graphicx}
\usepackage{subfigmat}
\usepackage{subfigure}
\usepackage{etoolbox}
\usepackage{graphicx,cite,epsfig,amssymb,amsmath,multirow}
\usepackage{graphicx,amsmath,amssymb,amsfonts,cite}
\usepackage{subfigure}
\usepackage{subfigmat}
\usepackage{etoolbox}
\usepackage{mathrsfs}
\usepackage{algorithm}
\usepackage{algorithmic}
\usepackage{float}
\usepackage{bm}
\usepackage{ulem}
\usepackage{color}
\usepackage{algorithm}
\usepackage{algorithmic}
\renewcommand{\algorithmicrequire}{\textbf{Input:}}
\renewcommand{\algorithmicensure}{\textbf{Output:}}

\makeatother

\begin{document}
\title{Enabling Covariance-Based Feedback in Massive MIMO: A User Classification Approach}
\author{\IEEEauthorblockN{Shuang Qiu\IEEEauthorrefmark{1},
David Gesbert\IEEEauthorrefmark{2}, and Tao Jiang\IEEEauthorrefmark{1}}
\IEEEauthorrefmark{1}{Huazhong University of Science and Technology, 430074 Wuhan, P. R. China}\\\IEEEauthorrefmark{2}{Communication systems, Eurecom, 06410 Biot, France}\\ Email: sqiu@hust.edu.cn, david.gesbert@eurecom.fr, Tao.Jiang@ieee.org}
\maketitle

\begin{abstract}
In this paper, we propose a novel channel feedback scheme for frequency division duplexing massive multi-input multi-output systems. The concept uses the notion of user statistical separability which was hinted in several prior works in the massive antenna regime but not fully exploited so far. We here propose a hybrid statistical-instantaneous feedback scheme based on a user classification mechanism where the classification metric derives from a rate bound analysis. According to classification results, a user either operates on a statistical feedback mode or instantaneous mode. Our results illustrate the sum rate advantages of our scheme under a global feedback overhead constraint.
\end{abstract}

\IEEEpeerreviewmaketitle

\section{Introduction}\label{sec_intro}
Massive multiple-input multiple-output (MIMO) is expected to be a key enabler for the next generation communication systems.
It has drawn considerable interest from  academia and industry for improving energy and spectral efficiency~\cite{ NgoApril2013,KongTWC2016}.

However, the large scale number of antennas brings up new challenges, one of which is the acquisition of accurate instantaneous channel state information (CSI), especially downlink CSI in frequency division duplex (FDD) massive MIMO systems.
In conventional MIMO FDD feedback mechanisms, users need to conduct channel estimation and feed the quantized channel back to base station (BS)~\cite{LoveJSAC2008} where quantization occurs over a certain budgeted number of bits.

Numerous works have been dedicated to reducing feedback overhead, such as trellis-extended codebook design~\cite{ChoiTWC2015}, compressive sensing-based channel feedback reduction~\cite{KuoWCNC2012}, channel covariance-based feedback representation~\cite{domeneTVT2015, WagnerTIT2012, GaoDaiTSP2015, AdhikaryOct.2013}.
In~\cite{AdhikaryOct.2013}, a two-stage precoding structure was introduced to reduce downlink pilot and channel feedback overhead.
%, in which channel covariance is used for user grouping and channel dimension reduction.
Advanced allocation of feedback bits among users was also proposed under total overhead constraint~\cite{brunoconf1}.
Furthermore, the channel feedback issue was tackled by exploiting user cooperation via device-to-device communications, such as~cooperative precoder feedback scheme~\cite{ChenYinTWC2017}.
While several previous schemes exploit the low rank behavior of channel covariance to map the instantaneous channel into a lower dimensional vector, the pair-wise orthogonality that may exist between users' signal subspaces \cite{AdhikaryOct.2013, YinGesbert2013} has not fully been exploited so far. This is because such orthogonality hinges on the degree of spatial separation between users and therefore cannot be guaranteed.

Still, we point out that, statistically speaking, some users are bound to be spatially isolated from the other users and can benefit the system by being identified and assigned a suitable feedback mechanism.
In fact, it is well known that a spatially isolated user can be served free of interference via a statistical precoder~\cite{YinGesbert2013, QiuTVT17}, hence not requiring any instantaneous feedback at all.

Therefore, this motivates a hybrid statistical-instantaneous feedback method where users are classified according to their channel covariance.
In this paper, some users are assigned zero bit towards instantaneous feedback and labeled as class-S users.
The remaining users share the feedback bit budget evenly and are named as class-I users.
The class-I users estimate their downlink instantaneous CSI and feed quantized channel back to BS.
A suitable precoder design is proposed for both user classes with statistical CSI of class-S users and instantaneous feedback of class-I users.
The precoder is based on signal-to-leakage-and-noise ratio (SLNR) and the closed-form precoding vectors are obtained.
Moreover, the user classification criterion is derived analytically from a sum rate bound maximization argument.
The proposed scheme is shown to improve significantly the overall system throughput in the presence of restricted feedback bit budget.

\section{ System and Channel Models}\label{sec_sys_mod}

We consider a single-cell massive MIMO system where the BS is equipped with $M$ antennas and serves~$K$ single-antenna users.
We assume the BS holds channel covariance of users.
When the proposed hybrid statistical-instantaneous
feedback method is considered, the~$K$ users are classified into two classes based on their channel covariance, i.e., $K_\mathrm{S}$ class-S and $K_\mathrm{I}$ class-I users.
%As shown in Fig.~\ref{system_model_conf},
Class-S users are assigned zero bit
towards instantaneous feedback, while class-I users share the feedback overhead evenly and feed their quantized instantaneous channel back to BS based on codebooks.

For any user $k$, a multipath channel model is considered~as
\vspace{-0.5em}
\begin{equation}\label{eq 10}
{\mathbf{h}_{k}} \triangleq \frac{1}{\sqrt{P}} \sum\limits_{p = 1}^P {\gamma _{{kp}}\mathbf{a} \left({\theta _{{kp}}}\right)}, k =1,\dots, K,
\end{equation}
where $P$ is the number of independent, identically distributed~(i.i.d.) paths,~$\gamma _{{kp}}$ represents the complex gain of the~$p$-th path and~$\mathbf{a} \left({\theta _{{kp}}}\right)$ is the corresponding steering vector.
For simplicity, uniform linear array is considered and the steering vector is given~as
\vspace{-0.5em}
\begin{equation}\label{eq 1111}
\mathbf{a}({\theta }_{{kp}} ) \triangleq  \left[ {\begin{array}{*{20}{c}}
1,\hspace{-0.7em}&{{e^{ j2\pi \frac{d}{\lambda }\sin ({\theta }_{{kp}} )}}} ,\hspace{-0.7em}& \cdots ,\hspace{-0.7em}& {{e^{ j2\pi \frac{{(M - 1)d}}{\lambda }\sin ({\theta }_{{kp}} )}}}
\end{array}} \hspace{-0.7em}\right]^T,
\end{equation}
where $d$ is antenna spacing, $\lambda$ is wavelength and ${\theta }_{{kp}}$ is the random angle of arrival (AoA) of the $p$-th path.

Throughout the paper, we use subscript $(\cdot)_{\mathrm{I},i}$ and $(\cdot)_{\mathrm{S},n}$ to denote the notations for the~$i$-th class-I and~$n$-th class-S users, respectively.
%Let~$K_\mathrm{S} $ and $K_\mathrm{I} $ denote the numbers of class-S and class-I users, respectively.
Then, the received signals $ {y_{\mathrm{I},i}}$ and ${y_{\mathrm{S},n}}$~are
\vspace{-0.5em}
\begin{align}
 {y_{\mathrm{I},i}} = p_d \mathbf{h}_{\mathrm{I},i}^{H}\left(\widehat{\mathbf{W}}_{\mathrm{I}} \mathbf{x}_{\mathrm{I}} + \widehat{\mathbf{W}}_{\mathrm{S}} \mathbf{x}_{\mathrm{S}}\right)+n_{\mathrm{I},i}, \label{eq 20170428.1}\\
 {y_{\mathrm{S},n}} =p_d \mathbf{h}_{\mathrm{S},n}^{H} \left(\widehat{\mathbf{W}}_{\mathrm{I}} \mathbf{x}_{\mathrm{I}} + \widehat{\mathbf{W}}_{\mathrm{S}} \mathbf{x}_{\mathrm{S}}\right) +n_{\mathrm{S},n},\label{eq 20170428.2}
\end{align}
respectively, where $ p_d$ is downlink transmit power,
%~$\mathbf{h}$ is downlink channel,~
$\widehat{\mathbf{W}}_\mathrm{I} \triangleq \left[\widehat{\mathbf{w}}_{\mathrm{I},1},\dots, \widehat{\mathbf{w}}_{\mathrm{I},K_\mathrm{I}}\right] \in \mathbb{C}^{M \times K_\mathrm{I}}$ and $\widehat{\mathbf{W}}_\mathrm{S} \triangleq \left[\widehat{\mathbf{w}}_{\mathrm{S},1},\dots, \widehat{\mathbf{w}}_{\mathrm{I},K_\mathrm{S}}\right] \in \mathbb{C}^{M \times K_\mathrm{S}}$ denote downlink precoding matrices with $\mathrm{E}\left\{ \left\|\mathbf{w}\right\|^2 \right\}=1$, $\mathbf{x}_{\mathrm{I}}$ and $\mathbf{x}_{\mathrm{S}}$ are consisted of i.i.d. downlink data~$x_{\mathrm{I},i}$ and~$x_{\mathrm{S},n}$ with~$\mathrm{E}\left\{ \mathbf{x} \mathbf{x}^H\right\}=\mathbf{I}$, respectively, and~$n$ denotes i.i.d. additive white Gaussian~noise (AWGN) with  zero mean and unit variance.
%Note that downlink precoding matrices $\widehat{\mathbf{W}}_\mathrm{I}$ and $\widehat{\mathbf{W}}_\mathrm{S}$ are designed based on the instantaneous feedback of class-I users and statistical CSI of class-S users.
We take class-I users as an example and~${y_{\mathrm{I},i}}$ is further given~as
\vspace{-0.6em}
\begin{equation}
\begin{split}\label{eq 5}
 {y_{\mathrm{I},i}}  =&
\underbrace {p_d \mathbf{h}_{\mathrm{I},i}^{H} {\widehat{\mathbf{w}}_{\mathrm{I},i}}{x_{\mathrm{I},i}}  }_{\textrm{Expected signal}} + \underbrace {p_d \sum\limits_{j = 1, j \ne i}^{K_{\mathrm{I}}  } {\mathbf{h}_{\mathrm{I},i}^{H} {\widehat{\mathbf{w}}_ {\mathrm{I},j}}{x_{\mathrm{I},j}}}}_{\textrm{ Interference from the other class-I users}}\\
& + \underbrace {p_d \sum\limits_{n = 1}^{K_{\mathrm{S}} } \mathbf{h}_{\mathrm{I},i}^{H}{\widehat{\mathbf{w}}_{\mathrm{S},n}}{{ {x}}_{\mathrm{S},n}}}_{\textrm{ Interference from class-S users}} +\underbrace {n_{\mathrm{I},i}}_{\textrm{AWGN}}.
\end{split}
\end{equation}
Then, the signal-to-interference-plus-noise ratio (SINR) of the~$i$-th class-I user is
\vspace{-0.4em}
\begin{align}
r_{\mathrm{I},i}&=\frac{{|{\mathbf{h}_{\mathrm{I},i}^H}{\widehat{\mathbf{w}}_{\mathrm{I},i}}{|^2}}} {{\sum\limits_{{{j = 1, j \ne i}} }^{K_{\mathrm{I}} } |{\mathbf{h}_{\mathrm{I},i}^H}{\widehat{\mathbf{w}}_{\mathrm{I},j}}{|^2} + \sum\limits_{n = 1}^{K_{\mathrm{S}}}  |{\mathbf{h}_{\mathrm{I},i}^H}{\widehat{\mathbf{w}}_{\mathrm{S},n}}{|^2} +  \frac{1}{p_d} }}.\label{sinr1}
%r_{\mathrm{S},n} &=\frac{{|{\mathbf{h}_{\mathrm{S},n}^H}{\widehat{\mathbf{w}}_{\mathrm{S},n}}{|^2}}} {{\sum\limits_{{{q = 1,q \ne n}} }^{K_{\mathrm{S}} } |{\mathbf{h}_{\mathrm{S},n}^H}{\widehat{\mathbf{w}}_{\mathrm{S},q}}{|^2} + \sum\limits_{i = 1}^{K_{\mathrm{I}}}  |{\mathbf{h}_{\mathrm{S},n}^H}{\widehat{\mathbf{w}}_{\mathrm{I},i}}{|^2} +  \frac{1}{p_d} }},\label{sinr2}
\end{align}
Similarly, we can obtain SINR $r_{\mathrm{S},n}$ for the $n$-th class-S user. Thus, the downlink ergodic sum rate of the system is
\vspace{-0.5em}
\begin{equation}\label{1.2.4}
R_{\mathrm{sum}} \hspace{-0.2em}= \hspace{-0.2em} \sum_{i=1}^{K_{\mathrm{I}}}\hspace{-0.2em} {\mathrm{E}\hspace{-0.2em} \left\{ {{{  \log }_2} \left( {1 \hspace{-0.2em}+\hspace{-0.2em} r_{\mathrm{I},i} } \right)} \right\}} + \sum_{n=1}^{K_{\mathrm{S}}} \hspace{-0.2em} {\mathrm{E}\left\{ {{{  \log }_2} \left( {1 \hspace{-0.2em} +\hspace{-0.2em} r_{\mathrm{S},n} } \right)} \right\}}.
\end{equation}

Note that, different from the systems under conventional feedback schemes, downlink precoding matrices $\widehat{\mathbf{W}}_\mathrm{I}$ and $\widehat{\mathbf{W}}_\mathrm{S}$ can only be designed based on the instantaneous feedback of class-I users and statistical CSI of class-S users.

\section{Downlink SLNR-Based Precoder Design }\label{sec_precoding}
Under the proposed hybrid statistical-instantaneous feedback scheme, a SLNR-based precoding method is proposed for both class-I and class-S users.
%\subsection{SLNR-Based Precoding Design}
Take the~$i$-th class-I user as an example, its SLNR is given~as
\vspace{-0.3em}
%\mathrm{E} \left\{  \right\}
\begin{align}\label{3.1.1}
%\begin{split}
\Gamma_{\mathrm{I},i}=\frac{ |\mathbf{h}_{\mathrm{I},i}^H \widehat{\mathbf{w}}_{\mathrm{I},i} |^2 }{ \sum\limits_{j=1,j\neq i}^{K_\mathrm{I}} | \mathbf{h}_{\mathrm{I},j}^H \widehat{\mathbf{w}}_{\mathrm{I},i}|^2 + \sum\limits_{n=1}^{{K_\mathrm{S}}}|\mathbf{h}_{\mathrm{S},n}^H \widehat{\mathbf{w}}_{\mathrm{I},i} |^2 +\frac{1}{p_d }  } .
\end{align}
Similarly, SLNR $\Gamma_{\mathrm{S},n}$ of the $n$-th class-S user can be obtained.
Since the instantaneous CSI of class-S users is unavailable at the BS, average SLNR $\mathrm{E} \left\{ \Gamma_{\mathrm{I},i} \right\}$ and~$\mathrm{E} \left\{ \Gamma_{\mathrm{S},n} \right\}$ are exploited~\cite{WangJinTSP2012}.
Moreover, based on the Mullen's inequality~$\mathrm{E} \left\{ \frac{X}{Y} \right\}\geq \frac{\mathrm{E} \left\{ X\right\}}{\mathrm{E} \left\{ Y \right\}}$, lower bounds of~$\mathrm{E} \left\{ \Gamma_{\mathrm{I},i} \right\}$ and~$\mathrm{E} \left\{ \Gamma_{\mathrm{S},n} \right\}$  are considered with the quantized channel of class-I users and channel covariance of class-S users as
\vspace{-0.2em}
\begin{align}
&\mathrm{E}\hspace{-0.3em} \left\{ \Gamma^{\mathrm{{LB}}}_{\mathrm{I},i} \right\} \hspace{-0.3em}=\hspace{-0.3em} \frac{  \widehat{\mathbf{w}}^H_{\mathrm{I},i}\overline{\mathbf{H}}_{\mathrm{I},i} \widehat{\mathbf{w}}_{\mathrm{I},i}}{\widehat{ \mathbf{w}}^H_{\mathrm{I},i}\hspace{-0.7em} \sum\limits_{{ {j = 1, j \ne i}}}^{K_{\mathrm{I}}} \hspace{-0.7em} \overline{\mathbf{H}}_{\mathrm{I},j}  \widehat{\mathbf{w}}_{\mathrm{I},i} +
\widehat{\mathbf{w}}^H_{\mathrm{I},i} \sum\limits_{n=1}^{K_{\mathrm{S}}} \mathbf{\Phi}_{\mathrm{S},n} \widehat{\mathbf{w}}_{\mathrm{I},i}+\hspace{-0.3em} \frac{1}{p_d}},\nonumber \\
%\end{align}%
%\begin{align}
&\mathrm{E}\hspace{-0.3em} \left\{ \Gamma^{\mathrm{{LB}}} _{\mathrm{S},n} \right\} \hspace{-0.3em}= \hspace{-0.3em} \frac{  \widehat{\mathbf{w}}^H_{\mathrm{S},n}\mathbf{\Phi}_{\mathrm{S},n} \widehat{\mathbf{w}}_{\mathrm{S},n}}{\widehat{\mathbf{w}}^H_{\mathrm{S},n} \hspace{-1em} \sum\limits_{{ {q = 1, q\ne n}}}^{K_{\mathrm{S}}}  \hspace{-0.5em} \mathbf{\Phi}_{\mathrm{S},q} \mathbf{h}_{\mathrm{S},n}^H  \widehat{\mathbf{w}}_{\mathrm{S},n} +
\widehat{\mathbf{w}}^H_{\mathrm{S},n} \hspace{-0.3em}  \sum\limits_{i=1}^{K_{\mathrm{I}}} \overline{\mathbf{H}}_{\mathrm{I},i} \widehat{\mathbf{w}}_{\mathrm{S},n}+\hspace{-0.3em} \frac{1}{p_d}},   \nonumber
%\end{split}
\end{align}
respectively, where $\overline{\mathbf{H}}_{\mathrm{I},i} \triangleq \widehat{\mathbf{h}}_{\mathrm{I},i} \widehat{\mathbf{h}}_{\mathrm{I},i}^H$ with $\widehat{\mathbf{h}}_{\mathrm{I},i}$ denoting the quantized channel of the $i$-th class-I user and $\mathbf{\Phi}_{\mathrm{S},n}$ is channel covariance of the $n$-th class-S user.
To maximize~$\mathrm{E} \left\{ \Gamma^{\mathrm{{LB}}}_{\mathrm{I},i} \right\}$ and~$\mathrm{E} \left\{ \Gamma^{\mathrm{{LB}}}_{\mathrm{S},n} \right\}$, the SLNR-based precoder is derived~as
\vspace{-0.7em}
\begin{align}
%\begin{split}
&\widehat{\mathbf{w}}_{\mathrm{I},i}\hspace{-0.3em}=\hspace{-0.3em} \mathbf{u}_\mathrm{max}\left\{ \hspace{-0.3em} \left(\sum\limits_{{ {j = 1, j \ne i}} }^{K_{\mathrm{I}}} \hspace{-0.5em} \overline{\mathbf{H}}_ {\mathrm{I},j}+\hspace{-0.3em} \sum\limits_{n=1}^{K_{\mathrm{S}}} \mathbf{\Phi}_{\mathrm{S},n} +\frac{\mathbf{I}_M}{p_d} \right)^{-1} \hspace{-1em} \overline{\mathbf{H}}_{\mathrm{I},i} \hspace{-0.3em} \right\}, \label{slnr1} \\
& \widehat{\mathbf{w}}_{\mathrm{S},n}\hspace{-0.3em} =\hspace{-0.3em} \mathbf{u}_\mathrm{max}\left\{  \hspace{-0.3em} \left( \sum\limits_{{{q = 1, q \ne n}} }^{K_{\mathrm{S}}}\hspace{-0.7em} \mathbf{\Phi}_{\mathrm{S},q}+ \hspace{-0.3em} \sum\limits_{i=1}^{K_{\mathrm{I}}} \overline{\mathbf{H}}_{\mathrm{I},i} + \frac{\mathbf{I}_M}{p_d}  \right)^{-1}  \hspace{-1.3em} \mathbf{\Phi}_{\mathrm{S},n} \hspace{-0.3em} \right\}, \label{slnr2}
%\end{split}
\end{align}
where $\mathbf{u}_\mathrm{max}(\mathbf{A})$ denotes the eigen-vector corresponding to the maximal eigenvalue of matrix $\mathbf{A}$.

\section{System Sum Rate Bound Analysis} \label{sec_sum_rate}
The system performance is highly impacted by the user classification due to the complex mutual interactions between users. Since only covariance matrices are available for user classification, we present a rate bound based on statistical information to predict the rate performance under the proposed SLNR-based precoder and any user classification solution.
\subsection{Approximate Subspace and Quantized Channel Prediction}\label{subsec_beam}

%The BS only holds channel covariance of users.
For user $k$, its channel covariance $ {{\mathbf{\Phi}}}_k$ can be equivalently represented in beam domain form $\widetilde{{\mathbf{\Phi}}}_k$~\cite{CaireTSP2017, SunJuneTC2015, XieApri2016} given~as
\vspace{-0.3em}
\begin{align}\label{eq ch3}
\widetilde{{\mathbf{\Phi}}}_k  = \mathbf{V } \widetilde{\mathbf{\Phi}}^{\mathrm{BD}}_k \mathbf{V } ^H,
\end{align}
where $\mathbf{V}$ is a  Discrete Fourier Transform (DFT) matrix whose~$t$-th column represents a virtual beam and is given as
\begin{equation}\label{eq ch2}
\hspace{-0.1em}\mathbf{v}\left( \frac{t}{M}\right) \triangleq \frac{1}{\sqrt{M}} \left[ {\begin{array}{*{20}{c}}
\hspace{-0.5em}1,\hspace{-1em}&{{e^{ j \pi (\frac{2t}{M}-1)}}} ,\hspace{-1em}& \cdots ,\hspace{-1em}& {{e^{ j \pi (M-1) (\frac{2t}{M}-1)}}}
\end{array}} \hspace{-0.5em} \right]^T.
\end{equation}
The matrix $\widetilde{\mathbf{\Phi}}^{\mathrm{BD}}_k$ is a diagonal matrix whose $t$-th diagonal element denotes the average gain of the $t$-th beam given as~$\mathrm{E} \left\{ \left| \left[{h}^{\mathrm{BD}}_{k}\right]_t \right|^2 \right\} $.
 %with~$\left| \left[ \widetilde{{h}}^{\mathrm{BD}}_{k}\right]_t\right|^2$ denoting the gain of the~$t$-th beam.
When $M \rightarrow \infty$, each path corresponds to only one virtual beam in $\mathbf{V}$ and the beam index $t$ satisfies $\left(\frac{2t}{M} \right)-1=\frac{2d}{\lambda}\sin\left( \theta_{kp}\right)$~\cite{ SunJuneTC2015}.
When $M$ is not large enough, power leakage exists.
But the most power concentrates around the closest beam~$\widetilde{{t}}=\lfloor t \rceil=\lfloor\frac{M}{2} \left(\sin(\theta_{kp})+1\right)\rceil$  with half
wavelength antenna spacing, where $\lfloor a \rceil$ denotes the closest integer to $a$.
Thus, the approximate complex gain of the~$\widetilde{t}$-th virtual beam can be obtained~as
%\begin{equation}\label{eq ch3}
$ \left|  [\widetilde{h}^{\mathrm{BD}}_ {k}\ ]_{\widetilde{t }} \right|^2 = \sum\limits_{p \in \mathcal{B}_{k,\widetilde{t}}} \left| \gamma _{{kp}} \right|^2$,
%\end{equation}
%Based on Lemma~1 given in \cite{XieTVT2017},
where~$\mathcal{B}_{k,\widetilde{t}}$ denotes the set of paths whose closest beam is the~$\widetilde{t}$-th beam.

The nonzero elements in~$\widetilde{\mathbf{\Phi}}^{\mathrm{BD}}_k$ are limited since the number of multipaths is usually small in massive MIMO systems.
We use~$d_{k,\mathrm{min}}$ and~$d_{k,\mathrm{max}}$ to denote the indices of the first and last nonzero elements, respectively.
Taking power leakage of paths to the neighbouring beams into consideration, we can present an approximate subspace of the $k$-th user as
\begin{equation}\label{eq eq2}
 \mathcal{S}_k =\mathrm{Span} \left\{\mathbf{v}\left(\frac{x}{M}\right) ,x_{k,\mathrm{min}}\leq x \leq x_{k,\mathrm{max}}  \right\},
\end{equation}
where $x_{k,\mathrm{min}} = \mathrm{max} (d_{k,\mathrm{min}}-x_k,1)$, $x_{k,\mathrm{max}} = \mathrm{min} (d_{k,\mathrm{max}}+x_k,M)$ and $x_k$ is a non-negative constant considered for power leakage, which can be obtained by empirical measurement.
%\footnote{The parameter $x_k$ is related with the number of BS antennas $M$ and SAoAs of users.
%Although it is difficult to obtain in theory analysis, it can be obtained from long-term statistics or off-line tables.
%When $x_k$ is zero, power leakage is not considered in the subspace expression.
%When $x_k>\mathrm{max }(M-d_{k,\mathrm{min}}, M-d_{k,\mathrm{max}})$, the power leakage is considered to spread in the whole~space. }.

The codebook design for spatially correlated channel usually takes channel covariance into account and codebook vectors are designed in subspaces.
Thus, a codebook~$\widetilde{\mathcal{C}}_k$ of size $X$ can be predicted with the assumption that codebook vectors are isotropically distributed in subspace~$ \mathcal{S}_k $.
Hence, the codebook vector $\widetilde{\mathbf{c}}_{k,u} \in \widetilde{\mathcal{C}}_k,  u=1,\dots, X,$ is constructed as %created~as
\vspace{-0.2em}
\begin{equation}\label{eq ch6}
\begin{split}
\widetilde{\mathbf{c}}_{k,u} & =
%\triangleq \mathbf{c}_k\left(\frac{u}{X'} \right)
%\\
  \frac{1}{\sqrt{M}} \left[ \hspace{-0.5em} {\begin{array}{*{20}{c}}
1,\hspace{-1em}&{{e^{ j \pi \theta_k(u)}}} ,\hspace{-1em}& \cdots ,& {{e^{ j \pi (M-1) \theta_k(u)}}}
\end{array}} \hspace{-0.5em} \right]^T,
\end{split}
\end{equation}
where~$\theta_k(u)=\left(\frac{2x_{k,\mathrm{min}}}{M}-1 \right)+ u \frac{2\left(x_{k,\mathrm{max}}-x_{k,\mathrm{min}} \right)}{MX}$.
%Thus, a codebook for spatially correlated channel is approximately presented in the form of DFT vectors.
%The codebook is

The class-I users choose the closest codebook vectors to their channel directions as quantized channels, such as $\widehat{\mathbf{h}}_{\mathrm{I},i}=\mathrm{arg} \mathop {{\mathop{\rm max}\nolimits} }\limits_{\mathbf{c}_u \in \widetilde{\mathcal{C}}_{\mathrm{I},i}} \; \left| \mathbf{h}_{\mathrm{I},i}^H \mathbf{c}_u \right|^2$.
However, instantaneous CSI of class-I users is unknown to the BS during user classification stage. Hence, the BS needs to predict channel directions based on channel covariance.
The idea is to take the beam $\mathbf{v}\left (\frac{t^*_{\mathrm{I},i}}{M}\right)$ in matrix~$\mathbf{V}$ with the largest average gain as predicted channel direction.
Then, the vector in codebook~$\widetilde{\mathcal{C}}_{\mathrm{I},i}$ which is closest to the vector $\mathbf{v}\left(\frac{t^*_{\mathrm{I},i}}{M}\right)$ is selected as feedback channel.
%, such as $\frac{2t^*_{\mathrm{I},i}}{M}-1=\theta_{\mathrm{I},i}(\widetilde{u}^*_ {{\mathrm{I},i}})$.
Thus, the predicted codebook vector index is obtained~as
\begin{equation}\label{eq feedback1}
\widetilde{u}^*_{{\mathrm{I},i}} = \left\{ {\begin{array}{*{20}{c}}
{\hspace{-5em}1,\hspace{7em} {t}^*_{\mathrm{I},i}= x_{{\mathrm{I},i},\mathrm{min}}}\\
\hspace{-0.5em}{\left\lfloor \frac{\left({t}^*_{\mathrm{I},i}- x_{{\mathrm{I},i},\mathrm{min}}\right)X} {x_{{{\mathrm{I},i},\mathrm{max}}}-x_{{\mathrm{I},i},\mathrm{min}}} \right\rceil, x_{{\mathrm{I},i},\mathrm{min}}< {t}^*_{\mathrm{I},i} \leq x_{{\mathrm{I},i},\mathrm{max}}} .
\end{array}} \right.
\end{equation}
Therefore, the BS obtains the predicted instantaneous channel feedback of class-I users.
%based on channel covariance and feedback~overhead.

\subsection{Lower Bound Analysis of System Sum Rate}

By substituting the predicted quantized channel and beam domain channel covariance $\widetilde{{\mathbf{\Phi}}}_k$ into (\ref{slnr1}) and (\ref{slnr2}), the SLNR-based precoding vectors for the~$n$-th class-S and the~$i$-th class-I users are approximately presented as
\vspace{-0.2em}
\begin{align}
 \widehat{\mathbf{w}}_{\mathrm{S},n}&=\mathbf{V} \mathbf{e}(\widetilde{l}_{\mathrm{S},n}^*), \label{slnr2.1}\\
 \widehat{\mathbf{w}}_{\mathrm{I},i}&=\mathbf{V} \mathbf{e}(\widetilde{\widehat{t}}_{\mathrm{I},i}), \label{slnr2.2}
\end{align}
respectively, where $\mathbf{e}(x)$ is the $x$-th column of an identity matrix, the index $\widetilde{l}_{\mathrm{S},n}^*$ is obtained by
%\begin{align}\label{eq lemma5}
${\widetilde{ {l}}^*_{\mathrm{S},n}} = \mathrm{arg}\mathop {\max }\limits_{l = 1, \ldots ,M}  \left[\widetilde{ {\mathbf{\Sigma}}}_{\mathrm{S},n} \right]_{l}$ and the $l$-th diagonal element is presented as
%\begin{equation}\label{eq ch22}
$ \left[ \widetilde{\mathbf{\Sigma}}_{\mathrm{S},n} \right]_{l}=\frac{ [ \widetilde{\mathbf{\Phi}}^{\mathrm{BD}}_{\mathrm{S},n} ]_l}{\sum\limits_{q=1,q\neq n}^{K_{\mathrm{S}}}  [\widetilde{\mathbf{\Phi}}^{\mathrm{BD}}_{\mathrm{S},q} ]_l+\sum\limits_{i=1}^{K_{\mathrm{I}}} \delta (\widetilde{\widehat{ {{t}}}}_{\mathrm{I},i}-l)+\frac{1}{p_d}}$ with ~$\delta(\cdot )$ the Kronecker delta function and $\widetilde{\widehat{ {{t}}}}_{\mathrm{I},i}=\left\lfloor x_{{\mathrm{I},i},\mathrm{min}}+ \frac{x_{{\mathrm{I},i},\mathrm{max}}-x_{{\mathrm{I},i},\mathrm{min}}}{X} \widetilde{u}^*_{\mathrm{I},i}\right\rceil$.
Due to lack of space, the detailed derivation are omitted.
%From Lemma \ref{lemma_slnr}, it is shown that the downlink precoding vectors can be predicted only based on the codebook vector indices $u^*_{\mathrm{I},i}$ and the beam domain channel covariance matrices of all the users.
%Therefore, the BS can easily predict the downlink precoding vector without feedback from class-I~users.

By substituting the approximate SLNR-based precoding given in (\ref{slnr2.1}) and~(\ref{slnr2.2}) into SINR $r_{\mathrm{I},i}$ and $r_{\mathrm{S},n}$, lower bounds $r^{\mathrm{LB}}_{\mathrm{I},i}$ and $r^{\mathrm{LB}}_{\mathrm{S},n}$ can be obtained.
In addition, we exploit effective SINR~$\mathrm{E }\left\{ {r}^{\mathrm{LB}}_{\mathrm{I},i} \right\}$ and~$\mathrm{E }\left\{ {r}^{\mathrm{LB}}_{\mathrm{S},n} \right\}$ to overcome the lack of class-S users' instantaneous CSI. Hence, a lower bound of sum rate is obtained as
\vspace{-0.2em}
\begin{align}\label{eq ch26}
\widetilde{{R}}^{\mathrm{LB}}_{\mathrm{sum}} &\hspace{-0.3em}=\hspace{-0.5em}\sum_{i=1}^{K_{\mathrm{I}}}  \hspace{-0.3em} \log\left(1+\mathrm{E }\left\{ {r}^{\mathrm{LB}}_{\mathrm{I},i} \right\} \right) \hspace{-0.3em} +\hspace{-0.5em} \sum_{n=1}^{K_{\mathrm{S}}}  \log\left(1+\mathrm{E }\left\{ {r}^{\mathrm{LB}}_{\mathrm{S},n}\right\} \right),
\end{align}
where the closed-form expressions of $\mathrm{E }\left\{ {r}^{\mathrm{LB}}_{\mathrm{I},i} \right\}$ and $\mathrm{E }\left\{ {r}^{\mathrm{LB}}_{\mathrm{S},n}\right\}$   are respectively given as
\vspace{-0.2em}
\begin{align}
\mathrm{E }\hspace{-0.3em} \left\{ {r}^{\mathrm{LB}}_{\mathrm{I},i} \right\} \hspace{-0.3em} =\hspace{-0.3em}\frac{ \mathrm{E}\left\{ \left|\left[ \widetilde{{h}}^{\mathrm{BD}} _{\mathrm{I},i}\right]_{\widetilde{\widehat{t}}_ {\mathrm{I},i}}  \right|^2 \right\}}{ \hspace{-0.5em} \sum\limits_{{{ {j=1, j\neq i}}} }^{K_{\mathrm{I}} } \hspace{-0.8em} \mathrm{E }\left\{\left|\left[ \widetilde{{h}}^{\mathrm{BD}} _{\mathrm{I},i}\right]_{\widetilde{\widehat{t}}_ {\mathrm{I},j}}  \right|^2\right\} \hspace{-0.3em}+ \hspace{-0.5em} \sum\limits_{n = 1}^{K_{\mathrm{S}}} \hspace{-0.3em} \mathrm{E }\left\{ \left|\left[ \widetilde{{h}}^{\mathrm{BD}} _{\mathrm{I},i}\right]_{\widetilde{l}^*_{\mathrm{S},n}}  \right|^2  \right\}\hspace{-0.3em}  +\hspace{-0.3em}  \frac{1}{p_d}  },\nonumber
\end{align}
\begin{align}
\mathrm{E }\hspace{-0.3em} \left\{ {r}^{\mathrm{LB}}_{\mathrm{S},n} \right\} \hspace{-0.3em}=\hspace{-0.3em} \frac{\mathrm{E }\left\{\left|\left[ \widetilde{ {h}}^{\mathrm{BD}} _{\mathrm{S},n}\right]_{\widetilde{l}^*_{\mathrm{S},n}}  \right|^2\right\}}{\hspace{-0.7em}  \sum\limits_{{ {q=1, q\neq n}}}^{K_{\mathrm{S}} } \hspace{-1em}  \mathrm{E }\left\{\left|\left[ \widetilde{ {h}}^{\mathrm{BD}} _{\mathrm{S},n}\right]_{\widetilde{l}^*_{\mathrm{S},q}}  \right|^2 \right\}\hspace{-0.3em} + \hspace{-0.3em}   \sum\limits_{i = 1}^{K_{\mathrm{I}}} \mathrm{E }\left\{\left|\left[ \widetilde{{h}}^{\mathrm{BD}} _{\mathrm{S},n}\right]_{\widetilde{\widehat{t}}_{\mathrm{I},i}}  \right|^2\right\}\hspace{-0.3em} +\hspace{-0.3em} \frac{1}{p_d}  }.\nonumber
\end{align}
%Thus, the approximate system sum rate can be directly calculated from beam domain channel covariance given the codebook size for each class-I user.
Therefore, given user classification and global feedback overhead budget~$B^{\mathrm{total}}$, a lower bound of system sum rate can be~calculated.

\subsection{ User Classification Algorithm for Sum Rate Maximization}
\label{sec_scheduling}
The optimal classifier for the proposed feedback scheme to maximize system sum rate is computationally complex.
Therefore, a sub-optimal user classifier is formulated based on the derived lower bound of sum rate and given as
\vspace{-0.3em}
\begin{subequations} \label{p2}
\begin{align}
\hspace{-1em}  \mathcal{K}^{\mathrm{sub}}_\mathrm{I}  ,  \mathcal{K}^{\mathrm{sub}}_\mathrm{S} & =  \arg \max  {\widetilde{R}^{\mathrm{LB}}_{\mathrm{sum}}}\left( {\mathcal{K} _\mathrm{I}} ,  {\mathcal{K} _\mathrm{S}} , B^{\mathrm{total}} \right) \label{q2.1}\\
\mathrm{s. t.} \;
K &= K_\mathrm{I}+K_\mathrm{S},\label{q2.3}\\
B & =\left\lfloor B^{\mathrm{total}}/K_{\mathrm{I}}\right\rfloor, \label{q2.4}
\end{align}
\end{subequations}
%Then, sub-optimal sets of class-I and class-S users can be obtained.
where ${\mathcal{K} _\mathrm{I}} $ and $  {\mathcal{K} _\mathrm{S}}$ denote the user sets of class-I and class-S users, respectively, $ \lfloor a \rfloor$ denotes the largest integer no more than $a$ and the constraint (\ref{q2.4}) denotes that class-I users share the total feedback overhead evenly and hence their codebook size is $X=2^B$.

To find the solution for problem~(\ref{p2}), a user classification algorithm is proposed in Alg. \ref{sum_rate_greedy}.
First, we assume all the~$K$ users are class-I users and calculate the lower bound of sum rate~$\widetilde{R}^{\mathrm{LB},K}_{\mathrm{sum}}$ based on (\ref{eq ch26}).
Then, we select the user who can achieve the largest $\widetilde{R}^{\mathrm{LB},K-1}_{\mathrm{sum}}$ as a new class-S user.
Repeat this procedure until all the users have been selected as class-S users.
Finally, compare all the $K+1$ sum rate~$\widetilde{R}^{\mathrm{LB},f}_{\mathrm{sum}}, f=0,\ldots,K,$ and select the maximal rate with index~$d^*$.
Thus, the optimal numbers of class-S and class-I users are~$K+1-d^*$ and~$d^*-1$, respectively.
Therefore, the optimal user set for class-S users consists of the first~$K+1-d^*$ selected class-S users and the remaining users are class-I users.

\begin{algorithm}[!t]
    \renewcommand{\algorithmicrequire}{\textbf{Input:}}
	\renewcommand{\algorithmicensure}{\textbf{Output:}}
	\caption{User Classification Algorithm with Sum Rate Bound Maximization}
    \label{sum_rate_greedy}
	\begin{algorithmic}[1]
        \REQUIRE %input
        $\widetilde{\mathbf{\Phi}}^{\mathrm{BD}}_k,k=1,\dots,K$, $B^{\mathrm{total}}$
		\ENSURE $ \mathcal{K}^{\mathrm{sub}}_\mathrm{I} $, $ \mathcal{K}^{\mathrm{sub}}_\mathrm{S} $
      			\STATE \textbf{Initialize} \\ Set $f=K$ and a rate vector $\widetilde{\mathbf{r}}_{\mathrm{sum}} =\emptyset$\\ The set of class-I users $ \mathcal{K}_\mathrm{I} =\{1,\dots,K  \}$\\
       The set of class-S users $ \mathcal{K}_\mathrm{S} =\emptyset$\\
       Calculate $\widetilde{R}^{\mathrm{LB},f}_{\mathrm{sum}}$ based on (\ref{eq ch26})\\
       Update $\widetilde{\mathbf{r}}_{\mathrm{sum}} =\left[\widetilde{\mathbf{r}}_{\mathrm{sum}} \; \widetilde{R}^{\mathrm{LB},f}_ {\mathrm{sum}}\right]$\\
\WHILE {$f\geq 0 $}
   \STATE Decrease $f$ by 1 and calculate $B=\left\lfloor  B^{\mathrm{total}}/f \right\rfloor$
   \STATE Find the user with index $n_\mathrm{S}$ as class-S user, such that
   \vspace{-0.5em}
\begin{equation}\label{eq ch30}
  n_\mathrm{S}=\mathrm{arg}\mathop {\max }\limits_{u \in   \mathcal{K}_\mathrm{I}  } \; \widetilde{R}^{\mathrm{LB},f}_{\mathrm{sum}} \hspace{-0.3em} \left(  \mathcal{K}_\mathrm{S} \cup \{u \},  \mathcal{K}_\mathrm{I}  \setminus \hspace{-0.3em} \{u\},B^{\mathrm{total}} \right) \nonumber
\end{equation}
   \STATE Update $ \mathcal{K}_\mathrm{S}  $ and $ \mathcal{K}_\mathrm{I} $ as
   \vspace{-0.5em}
\begin{equation}\label{eq alg1}
\begin{split}
 \mathcal{K}_\mathrm{S}   = \mathcal{K}_\mathrm{S} \cup \{n_\mathrm{S} \},
 \mathcal{K}_\mathrm{I} = \mathcal{K}_\mathrm{I} \setminus \{n_\mathrm{S} \} \nonumber
\end{split}
\end{equation}
   \STATE Update $\widetilde{\mathbf{r}}_{\mathrm{sum}}= \left[\widetilde{\mathbf{r}}_{\mathrm{sum}} \; \widetilde{R}^{\mathrm{LB},f}_{\mathrm{sum}} \right]$\\
\ENDWHILE
\STATE Find the maximal rate with index $d^*$ in vector $\widetilde{\mathbf{r}}_{\mathrm{sum}}$\\
\STATE The first $K+1-d^*$ users in $ \mathcal{K}_\mathrm{S} $ belong to~$ \mathcal{K}^{\mathrm{sub}}_\mathrm{S} $ and~$ \mathcal{K}^{\mathrm{sub}}_\mathrm{I} $ consists of the remaining users
\STATE \textbf{Return} $ \mathcal{K}^{\mathrm{sub}}_\mathrm{I} $, $ \mathcal{K}^{\mathrm{sub}}_\mathrm{S} $
    \end{algorithmic}
\end{algorithm}

\section{Simulation Results}\label{sec_simulation}

In this section, the analytical result and sum rate with the proposed hybrid statistical-instantaneous feedback scheme are evaluated.
%The proposed SLNR-based precoder and user classification mechanism are used.
%The parameter for subspace approximation is .
SLNR precoder~\cite{PatchCL2012} is used for conventional feedback scheme where all the users share the global feedback bit budget evenly and feed back quantized instantaneous channel.
Moreover, two types of codebook are considered:

\emph{1) DFT-based codebook}~\cite{YangHanzoICC2010}. The~$u$-th codebook vector is defined as~$\mathbf{c}_u \triangleq   \frac{1}{\sqrt{M}} \left[ \hspace{-0.5em} {\begin{array}{*{20}{c}}
1,\hspace{-1em}&{{e^{ j \pi (\frac{2u}{X}-1)}}} ,\hspace{-1em}& \cdots ,& {{e^{ j \pi (M-1) (\frac{2u}{X}-1)}}}
\end{array}} \hspace{-0.5em} \right]^T$.

\emph{2) Skewed codebook}~\cite{JiangCaireTWC2015}.
The codebook for the $i$-th class-I user is created as
%\begin{equation}\label{eq ch6}
$\mathcal{C}_{\mathrm{I},i}=\left\{ \mathbf{\Phi}^{1/2}_{\mathrm{I},i}\mathbf{f}_u / \left\| \mathbf{\Phi}^{1/2}_{{\mathrm{I},i}}\mathbf{f}_u \right\|\right\},$
%$\mathcal{C}_{k}=\left\{ \frac{\mathbf{\Phi}^{1/2}_{k}\mathbf{f}_u}{\left\| \mathbf{\Phi}^{1/2}_{k}\mathbf{f}_u \right\|}   \right\}$
%\end{equation}
where $\mathbf{f}_u, u=1,\dots,X,$ is isotropically distributed on the unit-sphere.
%This codebook design considers the channel covariance.
In the simulations, the AoA of the $P$ paths for user $k$ are uniformly distributed over~$\left[ \overline{\theta}_k -\theta_{\Delta}/2, \overline{\theta}_k +\theta_{\Delta}/2\right]$ with~$\overline{\theta}_k \in \left[ {-\frac{\pi}{2},\frac{\pi}{2}} \right]$ denoting the mean AoA and~$\theta_{\Delta}$ denoting spread AoA. We set~$P=20$ and~$\theta_\Delta = 10 ^\circ$. The parameter~$x_k$ for quantized channel prediction is 10.
The complex gain $\gamma_{kp}$ is with zero mean and unit variance.

\begin{figure}[!h]
\centering
\includegraphics[width=3.2in]{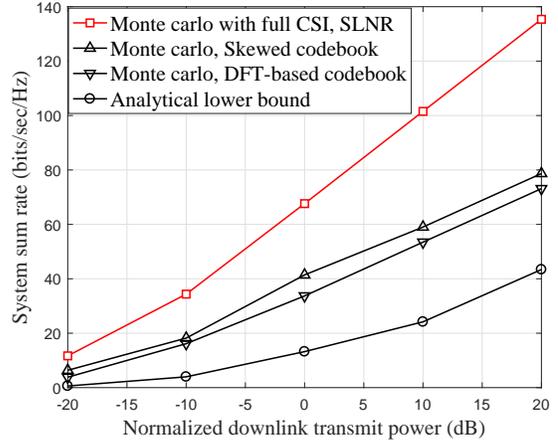}
\caption{ Performances comparison of Monte Carlo and the analytical lower bound results with $M=128$, $ K=10$ and $B^{\mathrm{total}}=40$ bits.}
\label{fig_simu_theory}
\end{figure}

Fig.~\ref{fig_simu_theory} depicts the system sum rate of the proposed feedback scheme with Monte Carlo and analytical lower bound derived in Section~\ref{sec_sum_rate}.
As a comparison, the Monte Carlo result of system sum rate with perfect downlink CSI is also provided.
Fig.~\ref{fig_simu_theory} shows that the analytical lower bound can reflect the change of system sum rate although it is derived only based on channel covariance and does not rely on codebook design.

\begin{figure}[!h]
\centering
\includegraphics[width=3.2in]{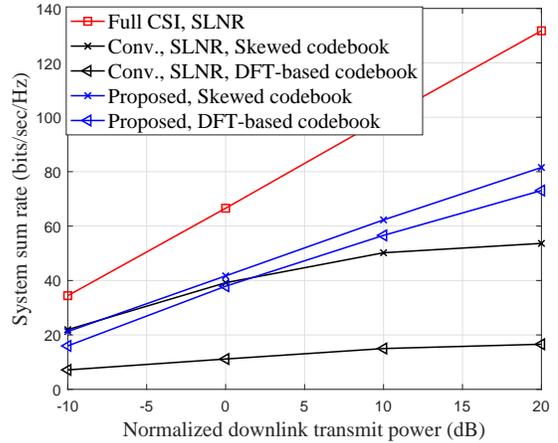}
\caption{System sum rate comparison versus downlink transmit power when $M=128$, $K=10$, $B^{\mathrm{total}}=40$ bits and $B_t=B^{\mathrm{total}}/K= 4$ bits for each user under conventional~scheme.}
\label{fig_snr_sumrate}
\end{figure}

The performance comparison of conventional and the proposed feedback schemes is provided in Fig.~\ref{fig_snr_sumrate}.
 %under different downlink transmit power.
It is shown that the proposed feedback scheme outperforms the conventional one, especially when DFT-based codebook is used.
The skewed codebook outperforms DFT-based codebook due to the consideration of channel statistics.
Moreover, the conventional scheme can only obtain marginal performance gain in high transmit power regime, while the performance of the proposed feedback scheme keeps rising with transmit power~increasing.

\begin{figure}[!h]
\centering
\includegraphics[width=3.2in]{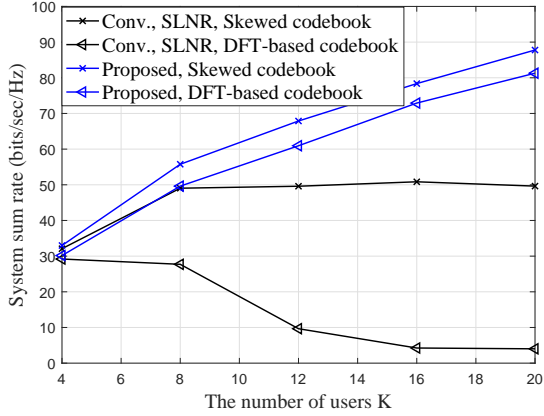}
\caption{System sum rate versus the number of users when $M=128$, $p_d=10$ dB, $B^{\mathrm{total}}=40$ bits and~$B_t =  B^{\mathrm{total}}/ K $ for each user under conventional~scheme.}
\label{fig_K_sumrate}
\end{figure}

It is shown from Fig.~\ref{fig_K_sumrate} that the performance of conventional scheme with DFT-based codebook deteriorates sharply with $K$ increasing, and the performance with skewed codebook is also restricted.
However, the performance of the proposed feedback scheme keeps increasing.
When $K=20$, the system sum rate under the proposed feedback scheme is 20 times larger than the conventional one with DFT-based codebook and 1.8 times larger with skewed codebook.
\begin{figure}[!h]
\centering
\includegraphics[width=3.2in]{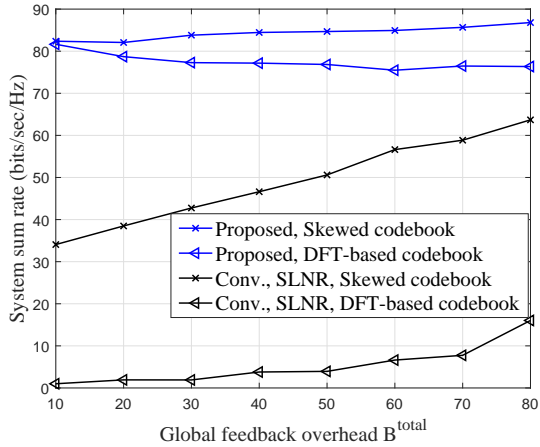}
\caption{System sum rate versus global feedback overhead when $M=128$, $K=20$, $p_d=10$ dB, $B_t = B^{\mathrm{total}}/K $ for each user under conventional~scheme.}
\label{fig_Btol_sumrate}
\end{figure}

Fig.~\ref{fig_Btol_sumrate} shows that the proposed feedback scheme achieves
much better performance than the conventional one, especially when total feedback bits are very limited, i.e., 10 feedback bits for 20 users.
With~$B^{\mathrm{total}}$ increasing, the performance of the proposed scheme with skewed codebook keep rising and the performance with DFT-based codebook slightly decreases.
%When total feedback bits are large enough, the performances of the conventional and the proposed feedback scheme will be identical.

\section{Conclusions}\label{sec_conclusion}
%In this paper, we propose a novel channel feedback scheme for frequency division duplexing massive multi-input multi-output systems. The concept uses the notion of user statistical separability which was hinted in several prior works in the massive antenna regime but not fully exploited so far. We here propose a hybrid statistical-instantaneous feedback scheme based on a user classification mechanism where the classification metric derives from a sum rate bound analysis. According to classification results, a user either operates on a statistical feedback mode or instantaneous mode. Our results illustrate the sum rate advantages of our scheme under a global feedback overhead constraint.

In this paper, a hybrid statistical-instantaneous feedback scheme was proposed, where only class-I users share the feedback bit
budget evenly and feed the quantized instantaneous channel back to BS.
We proposed a SLNR-based precoding method capable of handling the mixed statistical CSI of class-S users and quantized  instantaneous CSI of class-I users.
Then, we derived an analytical lower bound of system sum rate and proposed a user classification algorithm with the purpose of maximizing sum rate.
Finally, simulations verify the efficiency of the analytical result and illustrate that the novel feedback scheme significantly improves system sum rate even with very limited feedback bit budget.

\bibliographystyle{IEEEtran}
%\bibliography{Reference_FDD_feedback}

\end{document}